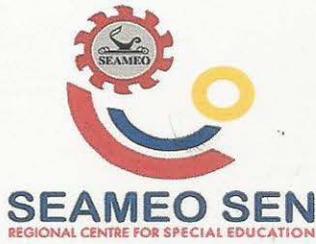 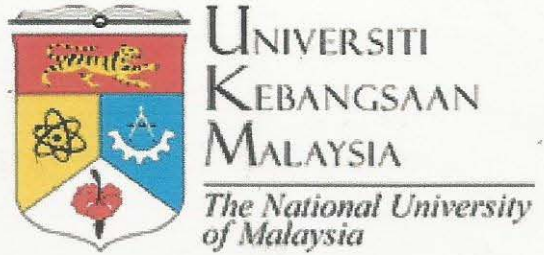 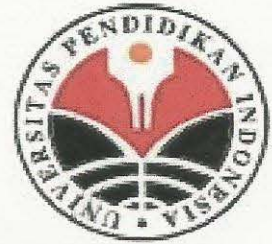

# INTERNATIONAL SEMINAR ON SPECIAL EDUCATION FOR SOUTHEAST ASIAN REGION

# &

# POSTGRADUATE SPECIAL EDUCATION UKM – UPI
# 4th SERIES 2014

*This Certificate is Awarded to*

**SALHAZAN NASUTION**

*As Presenter*

ONE STOP CAREER CENTRE FOR THE DISABLE PEOPLE

*Jointly Organized by*

**Faculty Of Education
Universiti Kebangsaan Malaysia**
with
**Universitas Pendidikan Indonesia**
and
**SEAMEO SEN**

*Date*
**25-28 January 2014**

*Venue*
**Faculty of Education
Universiti Kebangsaan Malaysia**

Dr. Yasmin Binti Hussain
Director
SEAMEO SEN

**Assoc. Prof. Dr. Mohd Hanafi Bin Mohd Yasin**
Chairman of the Seminar
Faculty of Education,
UKM, Bangi

Dr. Djadja Rahardja
Postgraduate School, UPI,
Indonesia

*Supported by*

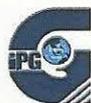 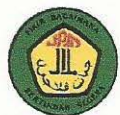 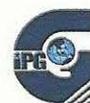

Jabatan Pendidikan Khas
IPGK Perempuan Melayu
Melaka

Jabatan Pendidikan
Negeri Melaka

IPG Kampus Pendidikan Teknik,
Bandar Enstek, Negeri Sembilan

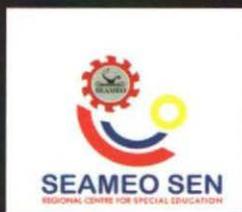
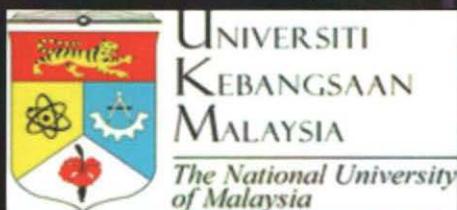
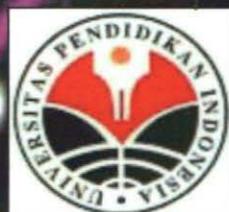

# PROCEEDINGS INTERNATIONAL SEMINAR OF POSTGRADUATE SPECIAL EDUCATION UKM - UPI- SEAMEO SEN 4th SERIES 2014

## IMPLEMENTATION OF INCLUSIVE EDUCATION FOR SPECIAL NEEDS CHILDREN

**Jointly Organized by**
FACULTY OF EDUCATION
UNIVERSITI KEBANGSAAN MALAYSIA
with
UNIVERSITAS PENDIDIKAN INDONESIA
and
SEAMEO SEN

**Editor**
Mohd Hanafi Mohd Yasin
Mohd Mokhtar Tahar
Safani Bari
Djadja Rahardja
Rozilawati Abdul Kadir

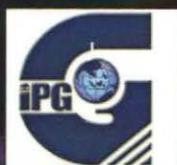
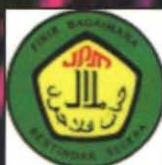
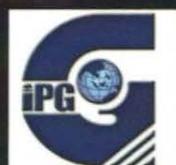

Jabatan Pendidikan Khas
IPG Kampus Perempuan Melayu Melaka

Jabatan Pendidikan Negeri Melaka

IPG Kampus Pendidikan Teknik, Bandar Enstek, Negeri Sembilan

# KATA PENGANTAR
## PENGERUSI SEMINAR INTERNASIONAL PASCASISWAZAH
## PENDIDIKAN KHAS UKM-UPI-SEAMEO SEN SIRI 4 2014
## UNIVERSITI KEBANGSAAN MALAYSIA

Assalamulaikum wbt.
Salam sejahtera dan selamat datang.

Pendidikan Khas telah melalui banyak perubahan dan kemajuan dalam sistem pendidikan di rantau Asia Tenggara. Sehingga kini Pendidikan Khas terus berkembang dan mengukuhkan kedudukannya dalam memberi pendidikan seterusnya menyediakan peluang dan ruang untuk golongan kurang upaya menyediakan diri untuk berdikari dan menyumbang kepada masyarakat dan negara. Namun begitu Pendidikan Khas tidak dinafikan masih memerlukan pelbagai penambahbaikan daripada pelbagai aspek. Sistem pendidikan yang lebih berwibawa dan dapat memenuhi keperluan golongan ini perlu diberi perhatian. Tanpa pendidikan yang sempurna, golongan kurang upaya akan ketinggalan dalam arus pembangunan yang sedang pesat ini. Justeru itu golongan kurang upaya harus diberikan hak pendidikan yang sama rata sebagaimana kanak-kanak biasa.

  Pendidikan inklusif kini sudah menjadi tuntutan dalam memberikan pendidikan kepada golongan kurang upaya. Kebanyakan negara maju di dunia sudah melaksanakannya. Pendidikan inklusif mempunyai banyak faedahnya kepada golongan kurang upaya. Selain dapat meningkatkan keyakinan diri, pendidikan inklusif juga boleh memberi kesedaran kepada masyarakat tentang kewujudan OKU. Melalui pendidikan inklusif masyarakat akan sedar tentang peranan mereka yang boleh dimainkan untuk memajukan OKU. Semua agensi dan institusi sebenarnya mempunyai tanggungjawab dalam meningkatkan pelaksanaan pendidikan inklusif. Pelaksanaan pendidikan inklusif yang berjaya dapat memberi impak yang positif kepada sistem pendidikan keseluruhannya. Semoga usaha murni ini menjadi pemangkin ke arah merealisasikan pendidikan inklusif sepenuhnya seperti mana yang dirangka dalam Pelan Pembangunan Pendidikan Malaysia (PPPM). Menurut PPPM, pendidikan inklusif harus sudah mencapai 30 peratus pelaksanaannya pada tahun 2015. Sehubungan itu, Seminar ini yang mengambil tema Pendidikan Inklusif sebagai tema utama akan menjadi satu landasan kepada seluruh intelektual, para pendidik dan ahli akademik dalam bidang Pendidikan Khas untuk bersama-sama bertukar-tukar fikiran, idea dan pandangan dalam meningkatkan pendidikan inklusif dan pendidikan khas.

  Semoga penerbitan prosiding ini akan menyediakan satu ruang ilmu yang bermakna kepada perkembangan pengajaran, pembelajaran, penyelidikan, perundingan dan latihan pendidikan inklusif dan pendidikan khas di semua peringkat sistem pendidikan. Seterusnya harapan yang tinggi diletakkan semoga siri-siri seminar ini akan terus menjadi medan perbincangan yang berterusan bagi semua pihak yang terlibat dengan pendidikan inklusif dan pendidikan khas.

Sekian dan terima kasih.
**Prof. Madya Dr. Mohd Hanafi Mohd Yasin**
Pengerusi Bersama, Seminar Internasional Pasca Siswazah Pendidikan Khas UKM-UPI-SEAMEO SEN Siri 4 2014

# ISI KANDUNGAN

# One Stop Career Centre for People with Disabilities

Salhazan Nasution [a], Mohd Hanafi Mohd Yasin [b1], Noraidah Sahari [c], Nor Syaidah Bahri [d],
Jamil Ahmad [e], Hasnah Toran [f], Safani Bari [g], Mohd Mokhtar Tahar [h]

[b,d,e,f,g,h] *Faculty of Education, Universiti Kebangsaan Malaysia, Bangi, Selangor, 43600, Malaysia*
[a,c] *Faculty of Information Science and Technology, Universiti Kebangsaan Malaysia, Bangi, Selangor, 43600, Malaysia*

## Abstract

Career is a journey of life that made the field of profession or employment options as a way to live. Careers are basis to generate income to sustain the needs of everyday life. Disabled people also need a job and benefit from the job same as a normal person. However, the attitudes of some prejudice community against the disabled people to find a job. The purpose of this paper is to discuss about website namely as one stop career centre for disabled people. This website helps disabled people who graduate from high school to find a job according to their qualifications and skills. Through this website, they can get information related to employment areas that they serve. Therefore, this website will help disabled people to get a job and to reduce the unemployment rate among them in Malaysia.

## 1. Introduction

Employment is essential and needed by human being, human need to work to meet their needs. The same applies to the disabled, they also need to work to fulfill their needs like a normal human. Generally, these people are not too weak to gets sympathy from others. Although they have certain disabilities, it does not mean that they are not able to do activities as a normal people. Although they are not able to do any work, but they are still able to use their mental and mind to make a contribution that can provide progress to this country, regardless of economic or social.

Disabled people have difficulty in obtaining productive employment and provide a source of income for their living. The number of disabled people who were employed in the public sector and the private sector is less encouraging. Labour Department in Malaysia data showed that the number of disable people working is lower compared to those who completed school (Madinah, 2012). In the public sector, government have been implemented namely a Service circular 10/1988 which provides the basis of taking 1% disabled configure each agency and the Appointing Authority (AA) must be disabled at least 1% of the number of officers in such agencies subject to the application by the disabled and the appropriateness of its functions and facilities (Persons with Disabilities Act, 2008). However, it does not guarantee disabled people will get 1% for every job that is offered. This is because the circular was still tied to the determination of eligibility and basic conditions. In addition, the provision of 1% either as an opportunity or actual placement is still not enough.

There are many challenges to be encountered by disabled people to enter the workforce. The challenges faced by them are a negative perception, fear and lack of confidence by the employer to take disabled people to work. In addition, there are also other issues that affect the opportunities for disabled people to enter the job market such as the issue of accessibility in and out of buildings (built environment), public places, workplaces and public transport. Awareness of employers, especially in the private sector is not much different from the past until now. Although some of the private sector employs the disabled people, it is only for lower positions as a sales assistant in a shopping mall cleaners etc. (Susan, 2012).

---

[1] Corresponding author. Tel.: +603-8921-6251; fax: +603-8925-4372.
*E-mail address*: mhmy6365@ukm.my

## 2. Problem Statement

Employment is an important asset in a man's life does not matter to normal people or people with disabilities. Employment is the main source of income and they are also responsible for providing and delivering various types of information. In addition, the ability to generate an employment workspace to create social ties between workers themselves and others. Those who do not have jobs will feel something down, especially financially, socially and psychologically. In other words, employment is a basic requirement for most people in the present age, where they think the job is a key determinant of quality of life (Zainida, 2009).

For disabled people, employment is a very important thing in their lives, but many people still ignore these things in the life of the disabled. People always impress that disabled people do not need a job to earn a better life like a normal person. They also argue that the aid money given is sufficient to accommodate disabled people's lives. In reality, disabled people are the most creative and productive if given the chance. Physical disability is not an obstacle to work and what's important, employers need to be rational with special offers to those involved. The advantage is that they are more focused than others. They also realized that if they work with half-heartedly, the opportunity to work will be destroyed and to get a new job is not easy. However, negative attitudes of employers who make the wrong impression on the ability of disabled people, such as the production of low productivity, high humility and high accident rate and barriers to employment.

Based on statistics from the Organization of the United Nations (2010), Malaysia has a population of approximately 28 million people. Based on these statistics, the number of disabled people in Malaysia is estimated at 2.8 million people. However, disabled people registered with the Social Welfare Department (SWD) of only 283 thousand. Heward (2000) found that there is a 50 % to 70% of people with disabilities who do not have jobs and among of them just received a minimum income. This is closely related to the attitude of some communities that show prejudice to the ability of the disabled to work. Attitudes of employers a still not opens their mind and are not ready to give disabled people the opportunity to work. Employer's refuse to give them jobs from several factors such as disabled workers need a lot of sick leave, higher insurance rates, safety in the workplace is disrupted, require special facilities and perhaps some changes had to be done in the workplace to ensure the safety of workers with disabilities. So, this will involve expenditure if employers employ them.

Although disable people receive an education like other normal people, opportunities for them to work are still limited. Thus, the current jobless rate among them a still high while pushing the system of education for the disabled reviewed (Ministry of Human Resources, 2006; Norani et al., 2005). Even facing with the competition in an increasingly challenging job market, disabled people should be trained in vocational skills which meet the requirements and needs. Until now, they remain to be fully accepted in the open job market.

The role played by each individual is different. Malaysia has a lot of job opportunities in various sectors such as agriculture, forestry, mining, electronics, wholesale and retail, manufacturing and others. However, opportunities for the disabled to find employment are limited. In general, suitable employment opportunities for disabled people in all sectors and occupations are still limited. Disabled people were not spared from contributing to further improvements in the political, economic, social, science and technology, and information technology. The high unemployment rate among disable people also gives them an impact to the social and economic status, self-image and self-esteem (Zainida, 2009).

## 3. Literature Review

The Department of Labour, Office of Disability Employment Policy (2001) in United Stated indicated that key goals of the one stop career centre approach, in its application to people with disabilities, were to streamline services, empower individuals, and provide universal accessibility. Streamlined services from all partners enabled activities and information to be co-located, coordinated, and integrated as a whole. Financial empowerment for individuals was achieved with Individual Training Accounts that allowed eligible adults to purchase training services, in conjunction with advice, guidance, and support through the one stop career centre system.

Other goals of the one stop career centre system were to develop an accurate performance assessment of its responsiveness to people with disabilities (Hall & Parker, 2005), and to create meaningful and seamless service delivery between the workforce and disability systems (Cohen, Timmons & Fesko, 2005). Seamless service delivery consists of a streamlined delivery of services by different agencies, and is smooth, coordinated, and efficient, reduces paperwork, avoids duplication, and links multiple programs into one system. Therefore, a person with a disability would not need to navigate multiple services systems in order to access a variety of necessary services, which can be overwhelming (Cohen et al., 2005). Instead, the partner agencies of the one stop career centre negotiate with each other to collaborate and coordinate their services, resulting in improved consumer outcomes and simplification of the process for customers.

If the one stop career centre system was successful in achieving the goal of increasing the employment rate of people with disabilities, economic statistics would reflect this, and would likewise, reflect an increase in their employment. Evaluation of the performance of the one stop career centre's employment services, however, revealed that Workforce Investment Act customers with disabilities were typically less likely to enter employment and retain employment when compared to peers without disabilities (Hall & Parker 2005). Holcomb & Barnow (2004) found that, although the program enrolled a number of people with disabilities, only a small proportion of them were actually served. Furthermore, people with disabilities who exited the program had lower employment and earnings than other existing customers (Holcomb & Barnow 2004).

In principle WIA's universal access provision along with the non-discrimination requirements of the Americans with Disabilities Act (ADA) addresses the accessibility needs of job seekers with disabilities at one stop. However, research shows that in practice persons with disabilities may still face a number of difficulties when utilizing the one stop system. Research has shown that not only the state workers, but also parents, teachers, and employers are not always aware of all the accessible resources for individuals with disabilities (Glazier & Tillmon, 1999). Some state workers may have more enhanced programs and educated employers while other state workers do not seem to know of all the programs and benefits available for individuals with disabilities.

Despite the efforts of the WIA's and the one stop career centre system, only about 37% of people with disabilities are reported in the national workforce (U.S. Census Bureau, 2006). Six per cent of adults aged 16 to 64 with a disability report the presence of a condition that makes it difficult to remain employed or to find a job (U.S. Census Bureau, 2006). The average earnings of workers with disabilities are lower; they are under-represented in the workforce, and experience both higher rates of poverty and limited access to employee benefits (Timmons 2002; U.S. Census Bureau, 2006). The Workforce Investment Act's Standardized Record Data revealed a significant decrease from 2001 to 2003 in percentages of people with a disability successfully exiting the one stop career centre program (National Council on Disability, 2005).

There was reluctance by the one stop career centre staff to serve people with disabilities because investment of the additional time and support required to assist those with disabilities could possibly result in inadequate outcomes to meet mandatory performance measures (Bader 2003; Hamner & Timmons 2005). It was found that employers who were aware of one stop career centres and used its services were more likely to be large and medium sized employers (U.S. Government Accountability Office, 2005).

## 4. One Stop Career Centre

Nowadays, with the amazing development and capabilities of information and communication technology is very help people with disabilities to find job. The ability of this technology is connecting people all over the world, store and disseminate information that can be accessed without going anywhere and done anywhere (Halimah, 2009). Thus, a career website for the disabled school graduates is being developed by researchers from the Faculty of Education, UKM.

The one stop career centre system was designed to integrate, collaborate, and upgrade relevant community programs and resources, and to provide employers with a larger pool of qualified, skilled workers (Rutgers, 2002). Federally mandated partnerships in one stop career centres provided individuals with meaningful and seamless access to information, services, and opportunities in the world of work. The one-stop career centre system has been described as a "no-wrong door" because it allows individuals with disabilities to choose, receives, and blends a variety of employment and training services through a single door (Dew, McGuire-Kuletz & Alan, 2001).

Productive and effective partnerships in one stop career centres also provide individuals with meaningful and seamless access to information, services, and opportunities in the world of work. One stop career centres are obligated to ensure that their facility and services are universally accessible to any individual seeking employment (Rutgers, 2002). So, people with disabilities have a right to use the one stop centre system, and are entitled to reasonable accommodations and modifications when using its services (Hoff, 2002). Unfortunately, many one stop career centres were not equipped to serve people with disabilities, and automatically referred these individuals to the public vocational rehabilitation system (Hoff, 2002).

In Malaysia, Ministry of Human Resources has introduced Placement System for Persons with disabilities (SPOKU), a system that was developed to help the disabled and those with special needs to find a job that suits their qualifications and abilities in line with the society that promoted by the government. After registration, the individual will be helped to find a job that fits their criteria and conditions. Not all disabled people can do the job like everyone else. An age, type of disability and no experience is the reasons why people are often rejected from job applications. In this system, employers can register vacancies reserved for disabled workers. As of March 2007, the system has already registered a total of 9,070 and has successfully placed a total of 6,799 people in the employment sector (Ministry of Human Resources, 2009).

Although SPOKU system was launched by the government to help people with disabilities to obtain employment, the percentage of those who use or access to the system is still not satisfactory because still not reached its target. Therefore, this study aimed to develop a new system of one stop career centre which can be accessed by disabled students for secondary school graduates to find a job. This system can also be accessed by teachers, school counsellors, parents and caregivers for the disabled to fill students' personal information into the system.

The cooperation of the teachers, parents and employers can play an important role for students with disabilities to help them make the transition from school to the working world. This website would be enhanced if more resources school-employer program is created, the involvement of teachers who are more aggressive, advanced training for students and the wider communication of a work between students and employers.

## 5. Conclusion

Disclosure of job opportunities for the disabled people can be done while they are still in school. Before they enter to job environment, the information related to the world of work for disabled can be collected by teachers and parents. In addition, they can get information at the school, internet, newspapers, radio, and television and so on. They can also get the information provided by government departments, NGOs, private agencies and etc. It is hoped that the development of this website will help the disabled people to get a job in line with their qualifications and skills. Indeed, all parties must play a role in opening up wider career opportunities to disabled so that they get the process to work smoothly and orderly. On the bright side, the cooperation from all parties allows for better employment opportunities for persons with disabilities. Giving a job to disabled in the private sector at least can reduce the unemployment rate of people with disabilities in Malaysia.

## References

Akta Orang Kurang Upaya. (2008). Retrieved January 20, 2013 from http://www.jkm.gov.my/images/stories/pdf/oku2008scan.pdf.

Bader, B.A. (2003). *Identification of best practices in one-stop career centers that facilitate use by people with disabilities seeking employment.* (Doctoral dissertation, Virginia Commonwealth University, Richmond). Retrieved January 9, 2013, from ProQuest Digital Dissertations database.

Cohen, A., Timmons, J.C. & Fesko, S.L. (2005). The Workforce Investment Act. How policy conflict and policy ambiguity affect implementation. *Journal of Disability Policy Studies, 15*(4), 221-230.

Dew, D.W., McGuire-Kuletz, M. & Alan, G.M. (Eds.). (2001). *Providing vocational rehabilitation in a workforce environment*. Retrieved October 1, 2012 from http://www.rcep6.org/IRI/IRI/27th_Workforce.pdf.

Glazier, R., & Tillmon, C. (1999). One-stop shopping and vocational rehabilitation, [Electronic version].*American Rehabilitation, 25*(3), 8-15.

Halimah Badiozezaman. (2009). Simbiosis Seni, Sains dan Teknologi Berasingan ke Multimedia-Fusion. Siri Syarahan Perdana UKM. Bangi: Penerbit UKM.

Hall, J.P. & Parker, K. (2005). One-stop career centers and job seekers with disabilities: Insights from Kansas. *Journal of Rehabilitation, 71*(4), 38-47. Retrieved October 2, 2012 from http://cwd.aphsa.org/publications/docs/One%20Stop%20Career%20Center%20and%20Job%20Seekers%20with%20Disabilities%20Insights%20from%20Kansas.pdf.

Hamner, D. & Timmons, J.C. (2005). *Case studies of local boards and one-stop centers. Underutilization of one-stops by people with significant disabilities*. Retrieved October 5, 2012 from http://www.communityinclusion.org/pdf/cs13.pdf.

Heward, W.L. (2000). *Exceptional Children: An Introduction to Special Education*. Upper Saddle River, NJ: Merrill.

Hoff, D. (2002). *One-stop career centres: Serving people with disabilities*. Retrieved September 20, 2012 from: http://www.onestops.info/article.php?article_id=69.

Holcomb, P. & Barnow, B.S. (2004). *Serving people with disabilities through the Workforce Investment Act's one-stop career centers*. Retrieved September 15, 2012 from http://www.urban.org/publications/411132.html.

Kementerian Sumber Manusia. (2013). Retrieved January 30, 2013 from http://jcs.mohr.gov.my/oku/.

Madinah Mohd Yusof, Mohd Hanafi Mohd Yasin, Siti Hawa Hashim & Mahidin Awang Itam. (2012). Transition programme and barriers to participating in the employment sector among hearing impaired students in Malaysia. *Procedia - Social and Behavioral Sciences* 47 (2012), 1793 – 180.

National Council on Disability. (2005). *NCD recommendations Workforce Investment Act reauthorization.* Retrieved December 30, 2012, from http://www.ncd.gov/newsroom/publications/2005/workforce_investment.htm.

Rutgers. (2002). *Workforce information customer satisfaction assessment: A primer for state and local planning*. Retrieved October 1, 2012 from http://www.heldrich.rutgers.edu/publications/workforce-information customersatisfaction-assessment-primer-state-and-local-planning.

Susan Binti Edwin. (2012). Faktor-faktor yang Mempengaruhi Kurangnya Pengambilan Golongan Orang Kelainan Upaya Bekerja di Sektor Swasta. Tesis Ijazah Sarjana Muda Teknologi Serta Pendidikan, Universiti Teknologi Malaysia.

Timmons, J.C., Fesko, S.L. & Cohen, A. (2004). Strategies of support: Increasing the capacity of one-stop centres to meet the needs of job seekers with disabilities. *Journal of Vocational Rehabilitation, 21*(1), 27-37.

Timmons, J.C., Schuster, J., Hamner, D. & Bose, J. (2002). Ingredients for success: Consumer perspectives on five essential elements to service delivery. *Journal of Vocational Rehabilitation*, *17*(3), 183-194.

U.S. Government Accountability Office. (2005). *Workforce Investment Act: Employers are aware of, using, and satisfied with one-stop services, but more data could help labor better address employers' needs. GAO-05-259.* Retrieved January 13, 2013, from http://www.gao.gov/new.items/d05259.pdf.

U.S. Census Bureau. (2006). *More than 50 million Americans report some level of disability*. Retrieved September 25, 2012, from http://www.census.gov/newsroom/releases/archives/aging_population/cb06-71.html.

Zainida Ariffin. (2009). Kerjaya untuk orang kurang upaya. Kuala Lumpur: PTS Professional Publishing Sdn.Bhd.